\documentclass{svproc}
\usepackage{makeidx}  
\usepackage{hyperref,graphicx}
\usepackage{amsmath,amssymb,mathrsfs}
\usepackage[utf8]{inputenc}
 
\newcommand{\beq}{\begin{equation}}
\newcommand{\eeq}{\end{equation}}
\newcommand{\bea}{\begin{eqnarray}}
\newcommand{\eea}{\end{eqnarray}}
\newcommand{\bit}{\begin{itemize}}
\newcommand{\eit}{\end{itemize}}
\newcommand{\ben}{\begin{enumerate}}
\newcommand{\een}{\end{enumerate}}

\usepackage{amssymb}

\makeindex
\usepackage{url}

\begin{document}
\frontmatter          
\pagestyle{headings}  
%

%
%
%
%

%
\title{Understanding Dwarf Galaxies in order to Understand Dark Matter}
\titlerunning{Dwarf Galaxies and Dark Matter}  
%
\author{Alyson M. Brooks}
\authorrunning{Alyson Brooks} 
%
\tocauthor{Alyson M. Brooks}
\institute{Rutgers, the State University of New Jersey  \\
Department of Physics \& Astronomy \\
136 Frelinghuysen Rd., Piscataway, NJ  08854\\
\email{abrooks@physics.rutgers.edu},\\ WWW home page:
\url{http://physics.rutgers.edu/~abrooks/index.html} \\
Submitted/reviewed as part of the Proceedings of the Simons Symposium on Illuminating Dark Matter}

\maketitle              

\begin{abstract}
Much progress has been made in recent years by the galaxy simulation community in making realistic galaxies, mostly by more accurately capturing the effects of baryons on the structural evolution of dark matter halos at high resolutions.  This progress has altered theoretical expectations for galaxy evolution within a Cold Dark Matter (CDM) model, reconciling many earlier discrepancies between theory and observations.  Despite this reconciliation, CDM may not be an accurate model for our Universe.  Much more work must be done to understand the predictions for galaxy formation within alternative dark matter models.
\end{abstract}
\section{Introduction: the need to understand baryons to understand dark matter}

Most of the matter in our Universe resides in an unknown component that we refer to as ``Dark Matter.'' There is six times more mass in dark matter than ordinary matter, which astronomers refer to as baryons. The large scale distribution of galaxies suggests that dark matter that is ``cold'' (because it travels slowly compared to the speed of light) provides an excellent description of our Universe \cite{Hlozek2012}.  However, when astronomers observe galaxies they are viewing only the ordinary matter that emits and absorbs photons.

Everything that we have learned about dark matter we have learned from astrophysics.\footnote{With one exception: dark matter direct detection experiments have ruled out a parameter space of cross-sections for interactions between dark matter and baryons.} Until a dark matter particle is detected, inferring the dark matter structure of galaxies is the primary method that we have right now to put constraints on {\it what} dark matter is.

For decades it has been assumed that, because there is so much more dark matter than ordinary matter, dark matter dominates the gravity in the Universe, and that wherever the dark matter is most dense, gas and stars must be there.  This assumption led theorists to make predictions for the formation of galaxies that either entirely neglected or poorly modeled the physics of gas and stars.  In doing so, a number of discrepancies between galaxy formation theory and observations were identified, particularly on ``small scales,'' i.e., in small galaxies and in the central regions of galaxies.  These discrepancies have evaded solution for so many years that they have become known collectively as the ``small scale crisis'' of the Cold Dark Matter (CDM) model for galaxy formation.

However, in the last few years there has been a paradigm shift, in which many astronomers now recognize the importance of including baryonic physics to solve CDM's small scale problems. Two of the most critical problems have been the ``cusp-core problem’’ and the ``missing satellites problem.’’ Both  problems are generally now agreed to be alleviated (or even solved) by the inclusion of baryonic physics.

Many simulators have demonstrated that energy injection from stars (usually referred to as ``feedback’’) in the form of both supernovae and energy from young, massive stars (i.e., ionization, radiation pressure, momentum injection from winds) can push the dark matter out of the central $\sim$kpc of galaxies by generating a repeated fluctuation in the potential well \cite{Read2005,deSouza2011,Pontzen2012,Teyssier2013}. This result reconciles the dark matter density profile predicted in CDM that is steeply rising toward the center (``cuspy'') \cite{Navarro1997,2008MNRAS.391.1685S,Navarro2010})
with observations  which instead prefer a shallower density slope or even a constant dark matter density ``core’' \cite{vandenbosch2000,deblok2001,deblok2002,Simon2003,Swaters2003,Weldrake2003,Kuzio2006,Gentile2007,Spano2008,Trachternach2008,deblok2008,Oh2011}.  Current simulations suggest that this process is most effective in dwarf galaxies with stellar masses $\sim$10$^8$ M$_{\odot}$ and halo virial masses of $\sim$10$^{10}$ M$_{\odot}$.  Below this mass, less star formation leads to less energy injection back to the interstellar gas in a galaxy, until there is simply not enough energy to alter the tightly bound cuspy dark matter profile.  At higher masses, the deeper potential wells of galaxies like the Milky Way seem to prevent core formation \cite{diCintio2014,Chan2015}.

Early simulations that included only dark matter found that there should be many more satellites that orbit around our Milky Way
galaxy within a CDM paradigm than we observe \cite{1999ApJ...524L..19M,1999ApJ...522...82K}.  Many of these satellites are expected
to be ``dark,'' unable to have formed stars due to photoevaporation of their gas when the Universe was re-ionized \cite{Quinn1996,Thoul1996,Barkana1999,Gnedin2000,Okamoto2008}, though this process alone cannot bring the predicted number of massive, luminous satellites into agreement with observation \cite{Brooks2013}.  
However, a simulation that includes baryons includes gas (by definition), which is able to cool itself (lose energy, primarily through radiation of photons).  This is in contrast to dark matter, which is unable to cool. Cooling gas adds more mass to the
center of the parent halo, creating stronger tidal forces that strip mass from satellite galaxies \cite{Penarrubia2010}, and can also destroy satellites that pass too near the disk. Thus, the presence of a disk (which doesn't exist in a dark matter-only simulation) brings both the numbers and kinematics of satellite galaxies into agreement with observations \cite{Zolotov2012,Brooks2013,Brooks2014,Wetzel2016,2017MNRAS.471.1709G}.

\section{Should you believe it?}

The importance of baryons in creating realistic galaxies and overcoming the small-scale problems of CDM is now recognized by many simulators.  Indeed, even simulators who do not have high enough resolution to resolve the processes that lead to dark matter core creation still find that inclusion of baryons can reconcile other outstanding challenges to CDM galaxy formation theory \cite{Sawala2016}. Thus, despite a range of star formation and feedback recipes, most simulators are now capable of simulating realistic galaxies that match a wide range of observed galaxy scaling relations (e.g., \cite{Brook2012,Aumer2013,Vogelsberger2013b}).  

\begin{figure*}[h]
\includegraphics[width=\textwidth]{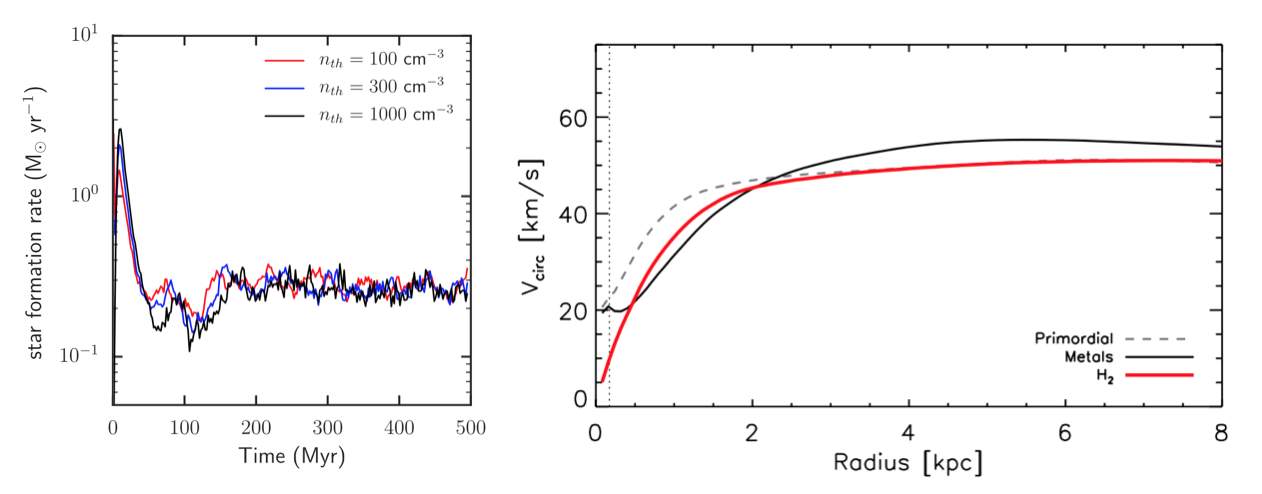}
\caption{{\it Left}: The SFR of a Milky Way-like simulated galaxy is unchanged as the density as which star formation is allowed to occur is changed (from 100 to 1000 amu cm$^{-3}$).  From \cite{Benincasa2016}.
{\it Right}: A dwarf galaxy simulated using three different prescriptions for star formation and feedback yields a similar result in all cases.  Shown here is the resulting rotation velocity.  From \cite{Christensen2014b}. }
\label{fig1brooks}
\end{figure*}

A common question from the non-simulators at the Simons Symposium was, ``What can be trusted in the simulations?’’  Much work has been done by simulators to address this same question.  Two of the key things that go into cosmological galaxy simulations but that vary most widely from simulation to simulation are the efficiency at which stars are formed and the form of the energy feedback.  A number of authors have now demonstrated that these two things are not independent; varying one will impact the other, with the net result being that galaxies converge to similar star formation rates and stellar masses because galaxies ``self-regulate,'' i.e., a change in the star formation is counter-balanced by subsequent feedback and vice versa \cite{saitoh08,Hopkins2011,Hopkins2013c,Christensen2014b,Benincasa2016,FIRE2,Pallottini2017}. Figure \ref{fig1brooks} shows results from two different investigations of this topic.  Self-regulation can occur as long as the resolution is high enough to capture the average densities in giant molecular clouds (GMCs), and therefore that the simulation is high enough resolution to have star formation limited to the scales of GMCs \cite{Agertz2016,Semenov2016}. Ref. \cite{Semenov2018} recently demonstrated that self-regulation is limited to the regime of strong feedback (which most of the highest resolution simulations fall under), which regulates the gas supply available to turn into stars. It is because of galaxy self-regulation that most simulators operating at high resolutions generally find similar results and come to similar conclusions about galaxy evolution, despite varying parameters.

Another common question from non-simulators was, ``What are the failings of the simulations?’’  This topic is always on the minds of simulators.  In general, the biggest question right now is whether galaxy simulations can reproduce the range of diverse galaxy rotation curves that are currently observed (see Manoj Kaplinghat’s summary in these proceedings).  This question applies across a range of galaxy mass scales.  At Milky Way masses, most simulations fail to create small stellar bulges, although the Milky Way and many of its largest spiral galaxy companions in the Local Volume seem to have small stellar bulges \cite{Kormendy2010,Shen2010,Fisher2011}.  Simulations that do create small stellar bulges in Milky Way-mass galaxies don’t seem to simultaneously be able to grow the disk as observed \cite{Aumer2014}.  On the other hand, most high resolution Milky Way-mass simulations do not currently include supermassive black holes with AGN feedback.  For a review on this topic, see Ref. \cite{Brooks2016}.

In smaller galaxies, simulators seem to be able to create diffuse dwarfs, but multiple authors have noted that they have not created compact dwarfs \cite{SantosSantos2018,Garrison-Kimmel2018}.  Possibly, this is due to small number statistics.  Because zoomed simulations are computationally expensive, the number of simulated galaxies is somewhat limited.  However, it is also possible that we have entered a phase in which feedback is too strong, preventing simulations from forming the densest and thinnest galaxies we observe (e.g., \cite{ElBadry2016}).  The current inability to reproduce the full range of diverse galaxies is being actively addressed amongst the simulation community.

\section{Implications for non-CDM models}

Finally, a common misunderstanding was identified by Symposium participants: despite the fact that baryons within a CDM model can reconcile theory with many observations, {\it this does not mean that alternative dark matter models are not worth pursuing}.  In fact, quite the opposite. A warm dark matter (WDM) model with baryons can still solve all of the small scale problems and remain consistent with observations, as can a self-interacting dark matter (SIDM) model with baryons.  There is no reason to believe that we understand all the properties of dark matter, and should therefore be pursuing a wide range of ideas. Thus, the question should really be: What are the predictions of alternative dark matter models with baryons included?

A sterile neutrino/dark fermion remains a viable dark matter candidate (see Kevork Abazajian’s contribution in this proceedings).  Some of the tightest constraints on WDM come from the abundance of low-mass satellite galaxies \cite{Maccio2010,Polisensky2011,Anderhalden2013,Horiuchi:2013noa} and the amount of small scale structure in the Lyman-$\alpha$ forest \cite{Viel2006,Seljak2006,Viel2008}.  In both cases, the WDM mass must be heavy enough that the data starts to look consistent with CDM, and the 3.5 keV line \cite{Boyarsky:2014jta,Bulbul:2014sua} that is possibly produced by decay of sterile neutrinos can still be made consistent with current observational constraints \cite{Abazajian:2014gza}.  The allowed mass range of a WDM particle is thus very tight.  Future x-ray telescopes should be able to resolve the 3.5 keV line, which will clarify its origin.  Additional constraints on WDM are likely to come from  the earliest epoch of star formation \cite{Barkana2001,Mesinger2005,deSouza2013,Pacucci2013,Governato2015,Chau2017}.  Because structure formation is delayed in WDM models, a delay of star formation with respect to CDM expectations may point to WDM as the correct model.

SIDM, on the other hand, is a model for which the constraints have only been loosening over the past few years.  After being initially invoked to solve the cusp-core problem \cite{Spergel:1999mh}, SIDM was quickly dismissed because it was believed to predict halo shapes that were more circular then observed \cite{Yoshida2000,Miralda2002}.  However, the question was revisited by Ref. \cite{Peter:2012jh}, who demonstrated that a cross section for interaction, $\sigma$, of about 1 cm$^2$/g (roughly the current limit in clusters, see Manoj Kaplinghat’s contribution in this proceedings) does not lead to enough change in the halo shapes of clusters to significantly distinguish them from CDM.  It has also been pointed out that the cross-section for interaction is likely to be velocity dependent, with particles moving slower relative to each other more likely to interact.  Ref. \cite{Loeb:2010gj} introduced such a model, allowing the constraints on cross-section at dwarf scales to be revisited.  

\begin{figure*}[h]
\includegraphics[width=\textwidth]{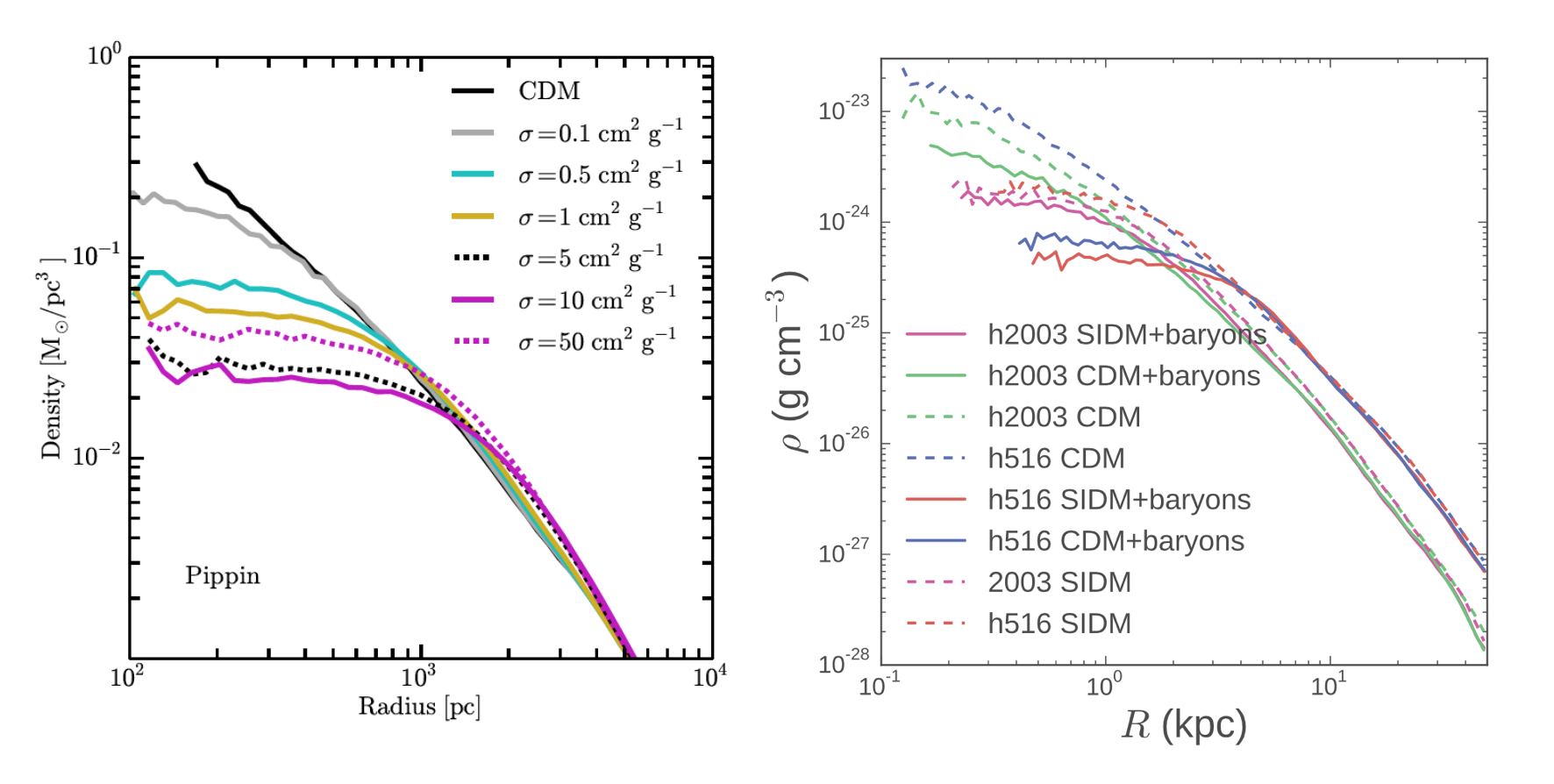}
\caption{{\it Left}: The density profiles for one dark matter halo of 9$\times$10$^9$ M$_{\odot}$ from Ref. \cite{Elbert:2014bma}. The results are similar enough for 0.5 $<$ $\sigma$ $<$ 50 cm$^2$/g that they would be observationally indistinguishable. 
{\it Right}: Simulation results from Ref. \cite{Fry2015}.  Two different dwarf galaxies are shown, run in varying models.  The h516 dwarf galaxy has a similar mass to the one run by Ref. \cite{Elbert:2014bma}.  Because baryons are effective at creating dark matter cores at this mass, and because baryonic core creation occurs before many SIDM scatterings in this 2 cm$^2$/g model, both the SIDM and CDM models with baryons result in nearly identical density profiles.  The h2003 dwarf, on the other hand, is low enough in stellar mass that baryonic effects  don't create a strong core, while the SIDM model creates more of a dark matter core.} 
\label{fig2brooks}
\end{figure*}

Ref. \cite{Elbert:2014bma} explored a dark matter-only SIDM simulation of a dwarf galaxy, at 9$\times$10$^9$ M$_{\odot}$ in halo mass.  This halo was resimulated with SIDM cross-sections of 0.1, 0.5, 1, 5, 10, and 50 cm$^2$/g. Figure \ref{fig2brooks} shows the resulting dark matter density profiles.  In essence, they are all similar enough that it would be an observational challenge to try to distinguish between results from 0.5 to 50 cm$^2$/g.  The 50 cm$^2$/g model is currently the largest cross-section explored to date. Even this large cross-section cannot be ruled out, with it’s density profile being comparable to what is inferred in observed dwarf galaxies.

Ref. \cite{Elbert:2014bma} did not include baryonic physics, and the picture is altered further when baryons are considered.  Ref. \cite{Fry2015} simulated a dwarf galaxy of comparable mass to the one run by \cite{Elbert:2014bma}, but with baryons.  Ref. \cite{Fry2015} ran two models: an SIDM model with a cross-section of 2 cm$^2$/g, and a standard CDM model.  They discovered that the baryons begin to form a core before the SIDM model has enough time to start significantly scattering particles to create a core.  Because of this, the resulting simulated dwarfs were identical (see Figure \ref{fig2brooks} for their density profiles).  

The largest cross-section yet explored with baryons in this same galaxy mass range is $\sim$20 cm$^2$/g (in the vdSIDMa simulation in Ref. \cite{Vogelsberger2014}), and the results were entirely consistent with observations.  Thus, there are currently no real constraints on the largest allowed cross-section at dwarf galaxy scales.  To constrain the particle physics models, two approaches should be taken: an observational approach and a simulation approach. 

Observationally, there are a few hints that should be pursued further.  First, in SIDM there are regimes where the baryons are likely to follow the dark matter distribution, e.g., in dark matter-dominated dwarf galaxies when the cross-section is large \cite{Vogelsberger2014}.  The extent to which baryons trace DM needs to be explored in more detail, across a range of dwarf galaxy masses (from ultra-faints up to LMC-mass galaxies) and cross-sections.  Related, Ref. \cite{Dooley2016} found that SIDM satellites that fall into the Milky Way have their stellar orbits expanded as the halos get tidally stripped.  Can the sizes of observed satellites be used to point to a DM model?

Second, it has been noted that baryons do not provide enough energy to create cores in the ultra-faint dwarf galaxy range, with stellar masses $<$ 10$^5$ M$_{\odot}$.  However, SIDM could create cores in these small halos (see comparison of h2003 SIDM and CDM models in Figure \ref{fig2brooks}).  Thus, measurements of the central densities and density slopes in the inner regions of dwarf galaxies are essential.  Unfortunately, determining the central densities is a challenge.  Even when scientists use the same data set, they have come to different conclusions about the presence of cores in dwarf galaxies (e.g., \cite{Breddels2013,Walker2011,Strigari2014}) while ultra-faints contain many fewer stars and will thus be even more of a challenge.  However, tackling the problem of how to determine central densities in dwarfs is an absolute priority in determining properties of dark matter — it could very well lead to a ``smoking gun’’ that identifies or rules out a dark matter model.

From a simulation perspective, the ideal approach would be to crank up the SIDM interaction cross-section and ask when galaxy formation breaks.  That is, when do the simulation results stop being consistent with observations?  A range of masses, from ultra-faint dwarf galaxies to the classical dwarf mass scale used above, should be investigated.  However, resources are limited, both in terms of computing time and human resources.  Individual simulations can require millions of CPU hours (and to fully explore SIDM, likely up to 100 million CPU hours would be needed).  Meanwhile, few people are working on this topic given that the field is biased toward a preference for CDM with baryons.  

\section{Summary}

In summary, there has been much progress in understanding the role of baryons in galaxy evolution in the past decade. To much of the astronomy community, this has solidified their confidence in the CDM model.  However, it should primarily solidify their confidence in the ability of simulations to model baryonic physics.  We still do not understand dark matter, and our favorite WIMP model continues to elude detection.  Thus, we need to have an open mind about the possible properties of dark matter.  A wide range of properties still waits to be explored in terms of consistency with galaxy evolution.

\bigskip
\bigskip
\noindent {\bf Acknowledgements} Thank you to the Simons Foundation for hosting this Symposium, and to the organizers for bringing together a truly stimulating group of dark matter scientists.  My work on baryons within a CDM model has been funded by NSF awards AST- 1411399 and AST-1813871, and by the Space Telescope Science Institute awards HST-AR-13925 and HST-AR-14281.



\newpage
\bibliography{simons-symp-refs}

\begin{thebibliography}{10}
\providecommand{\url}[1]{{#1}}
\providecommand{\urlprefix}{URL }
\expandafter\ifx\csname urlstyle\endcsname\relax
  \providecommand{\doi}[1]{DOI \discretionary{}{}{}#1}\else
  \providecommand{\doi}{DOI \discretionary{}{}{}\begingroup
  \urlstyle{rm}\Url}\fi

\bibitem{Hlozek2012}
R.~{Hlozek}, J.~{Dunkley}, G.~{Addison}, J.W. {Appel}, J.R. {Bond}, C.~{Sofia
  Carvalho}, S.~{Das}, M.J. {Devlin}, R.~{D{\"u}nner}, T.~{Essinger-Hileman},
  J.W. {Fowler}, P.~{Gallardo}, A.~{Hajian}, M.~{Halpern}, M.~{Hasselfield},
  M.~{Hilton}, A.D. {Hincks}, J.P. {Hughes}, K.D. {Irwin}, J.~{Klein},
  A.~{Kosowsky}, T.A. {Marriage}, D.~{Marsden}, F.~{Menanteau}, K.~{Moodley},
  M.D. {Niemack}, M.R. {Nolta}, L.A. {Page}, L.~{Parker}, B.~{Partridge},
  F.~{Rojas}, N.~{Sehgal}, B.~{Sherwin}, J.~{Sievers}, D.N. {Spergel}, S.T.
  {Staggs}, D.S. {Swetz}, E.R. {Switzer}, R.~{Thornton}, E.~{Wollack}, \apj
  \textbf{749}, 90 (2012).
\newblock \doi{10.1088/0004-637X/749/1/90}

\bibitem{Read2005}
J.I. {Read}, G.~{Gilmore}, \mnras \textbf{356}, 107 (2005).
\newblock \doi{10.1111/j.1365-2966.2004.08424.x}

\bibitem{deSouza2011}
R.S. {de Souza}, L.F.S. {Rodrigues}, E.E.O. {Ishida}, R.~{Opher}, \mnras
  \textbf{415}, 2969 (2011).
\newblock \doi{10.1111/j.1365-2966.2011.18916.x}

\bibitem{Pontzen2012}
A.~{Pontzen}, F.~{Governato}, \mnras \textbf{421}, 3464 (2012).
\newblock \doi{10.1111/j.1365-2966.2012.20571.x}

\bibitem{Teyssier2013}
R.~{Teyssier}, A.~{Pontzen}, Y.~{Dubois}, J.I. {Read}, \mnras \textbf{429},
  3068 (2013).
\newblock \doi{10.1093/mnras/sts563}

\bibitem{Navarro1997}
J.F. {Navarro}, C.S. {Frenk}, S.D.M. {White}, \apj \textbf{490}, 493 (1997).
\newblock \doi{10.1086/304888}

\bibitem{2008MNRAS.391.1685S}
V.~{Springel}, J.~{Wang}, M.~{Vogelsberger}, A.~{Ludlow}, A.~{Jenkins},
  A.~{Helmi}, J.F. {Navarro}, C.S. {Frenk}, S.D.M. {White}, \mnras
  \textbf{391}, 1685 (2008).
\newblock \doi{10.1111/j.1365-2966.2008.14066.x}

\bibitem{Navarro2010}
J.F. {Navarro}, A.~{Ludlow}, V.~{Springel}, J.~{Wang}, M.~{Vogelsberger},
  S.D.M. {White}, A.~{Jenkins}, C.S. {Frenk}, A.~{Helmi}, \mnras \textbf{402},
  21 (2010).
\newblock \doi{10.1111/j.1365-2966.2009.15878.x}

\bibitem{vandenbosch2000}
F.C. {van den Bosch}, B.E. {Robertson}, J.J. {Dalcanton}, W.J.G. {de Blok}, \aj
  \textbf{119}, 1579 (2000).
\newblock \doi{10.1086/301315}

\bibitem{deblok2001}
W.J.G. {de Blok}, S.S. {McGaugh}, V.C. {Rubin}, \aj \textbf{122}, 2396 (2001).
\newblock \doi{10.1086/323450}

\bibitem{deblok2002}
W.J.G. {de Blok}, A.~{Bosma}, \aap \textbf{385}, 816 (2002).
\newblock \doi{10.1051/0004-6361:20020080}

\bibitem{Simon2003}
J.D. {Simon}, A.D. {Bolatto}, A.~{Leroy}, L.~{Blitz}, \apj \textbf{596}, 957
  (2003).
\newblock \doi{10.1086/378200}

\bibitem{Swaters2003}
R.A. {Swaters}, B.F. {Madore}, F.C. {van den Bosch}, M.~{Balcells}, \apj
  \textbf{583}, 732 (2003).
\newblock \doi{10.1086/345426}

\bibitem{Weldrake2003}
D.T.F. {Weldrake}, W.J.G. {de Blok}, F.~{Walter}, \mnras \textbf{340}, 12
  (2003).
\newblock \doi{10.1046/j.1365-8711.2003.06170.x}

\bibitem{Kuzio2006}
R.~{Kuzio de Naray}, S.S. {McGaugh}, W.J.G. {de Blok}, A.~{Bosma}, \apjs
  \textbf{165}, 461 (2006).
\newblock \doi{10.1086/505345}

\bibitem{Gentile2007}
G.~{Gentile}, P.~{Salucci}, U.~{Klein}, G.L. {Granato}, \mnras \textbf{375},
  199 (2007).
\newblock \doi{10.1111/j.1365-2966.2006.11283.x}

\bibitem{Spano2008}
M.~{Spano}, M.~{Marcelin}, P.~{Amram}, C.~{Carignan}, B.~{Epinat},
  O.~{Hernandez}, \mnras \textbf{383}, 297 (2008).
\newblock \doi{10.1111/j.1365-2966.2007.12545.x}

\bibitem{Trachternach2008}
C.~{Trachternach}, W.J.G. {de Blok}, F.~{Walter}, E.~{Brinks}, R.C.
  {Kennicutt}, Jr., \aj \textbf{136}, 2720 (2008).
\newblock \doi{10.1088/0004-6256/136/6/2720}

\bibitem{deblok2008}
W.J.G. {de Blok}, F.~{Walter}, E.~{Brinks}, C.~{Trachternach}, S.~{Oh}, R.C.
  {Kennicutt}, \aj \textbf{136}, 2648 (2008).
\newblock \doi{10.1088/0004-6256/136/6/2648}

\bibitem{Oh2011}
S.H. {Oh}, C.~{Brook}, F.~{Governato}, E.~{Brinks}, L.~{Mayer}, W.J.G. {de
  Blok}, A.~{Brooks}, F.~{Walter}, \aj \textbf{142}, 24 (2011).
\newblock \doi{10.1088/0004-6256/142/1/24}

\bibitem{diCintio2014}
A.~{Di Cintio}, C.B. {Brook}, A.V. {Macci{\`o}}, G.S. {Stinson}, A.~{Knebe},
  A.A. {Dutton}, J.~{Wadsley}, \mnras \textbf{437}, 415 (2014).
\newblock \doi{10.1093/mnras/stt1891}

\bibitem{Chan2015}
T.K. {Chan}, D.~{Kere{\v s}}, J.~{O{\~n}orbe}, P.F. {Hopkins}, A.L. {Muratov},
  C.A. {Faucher-Gigu{\`e}re}, E.~{Quataert}, \mnras \textbf{454}, 2981 (2015).
\newblock \doi{10.1093/mnras/stv2165}

\bibitem{1999ApJ...524L..19M}
B.~{Moore}, S.~{Ghigna}, F.~{Governato}, G.~{Lake}, T.~{Quinn}, J.~{Stadel},
  P.~{Tozzi}, \apjl \textbf{524}, L19 (1999).
\newblock \doi{10.1086/312287}

\bibitem{1999ApJ...522...82K}
A.~{Klypin}, A.V. {Kravtsov}, O.~{Valenzuela}, F.~{Prada}, \apj \textbf{522},
  82 (1999).
\newblock \doi{10.1086/307643}

\bibitem{Quinn1996}
T.~{Quinn}, N.~{Katz}, G.~{Efstathiou}, \mnras \textbf{278}, L49 (1996)

\bibitem{Thoul1996}
A.A. {Thoul}, D.H. {Weinberg}, \apj \textbf{465}, 608 (1996).
\newblock \doi{10.1086/177446}

\bibitem{Barkana1999}
R.~{Barkana}, A.~{Loeb}, \apj \textbf{523}, 54 (1999).
\newblock \doi{10.1086/307724}

\bibitem{Gnedin2000}
N.Y. {Gnedin}, \apj \textbf{542}, 535 (2000).
\newblock \doi{10.1086/317042}

\bibitem{Okamoto2008}
T.~{Okamoto}, L.~{Gao}, T.~{Theuns}, \mnras \textbf{390}, 920 (2008).
\newblock \doi{10.1111/j.1365-2966.2008.13830.x}

\bibitem{Brooks2013}
A.M. {Brooks}, M.~{Kuhlen}, A.~{Zolotov}, D.~{Hooper}, \apj \textbf{765}, 22
  (2013).
\newblock \doi{10.1088/0004-637X/765/1/22}

\bibitem{Penarrubia2010}
J.~{Pe{\~n}arrubia}, A.J. {Benson}, M.G. {Walker}, G.~{Gilmore}, A.W.
  {McConnachie}, L.~{Mayer}, \mnras \textbf{406}, 1290 (2010).
\newblock \doi{10.1111/j.1365-2966.2010.16762.x}

\bibitem{Zolotov2012}
A.~{Zolotov}, A.M. {Brooks}, B.~{Willman}, F.~{Governato}, A.~{Pontzen},
  C.~{Christensen}, A.~{Dekel}, T.~{Quinn}, S.~{Shen}, J.~{Wadsley}, \apj
  \textbf{761}, 71 (2012)

\bibitem{Brooks2014}
A.M. {Brooks}, A.~{Zolotov}, \apj \textbf{786} (2014)

\bibitem{Wetzel2016}
A.R. {Wetzel}, P.F. {Hopkins}, J.h. {Kim}, C.A. {Faucher-Gigu{\`e}re},
  D.~{Kere{\v s}}, E.~{Quataert}, \apjl \textbf{827}, L23 (2016).
\newblock \doi{10.3847/2041-8205/827/2/L23}

\bibitem{2017MNRAS.471.1709G}
S.~{Garrison-Kimmel}, A.~{Wetzel}, J.S. {Bullock}, P.F. {Hopkins},
  M.~{Boylan-Kolchin}, C.A. {Faucher-Gigu{\`e}re}, D.~{Kere{\v s}},
  E.~{Quataert}, R.E. {Sanderson}, A.S. {Graus}, T.~{Kelley}, \mnras
  \textbf{471}, 1709 (2017).
\newblock \doi{10.1093/mnras/stx1710}

\bibitem{Sawala2016}
T.~{Sawala}, C.S. {Frenk}, A.~{Fattahi}, J.F. {Navarro}, R.G. {Bower}, R.A.
  {Crain}, C.~{Dalla Vecchia}, M.~{Furlong}, J.C. {Helly}, A.~{Jenkins}, K.A.
  {Oman}, M.~{Schaller}, J.~{Schaye}, T.~{Theuns}, J.~{Trayford}, S.D.M.
  {White}, \mnras \textbf{457}, 1931 (2016).
\newblock \doi{10.1093/mnras/stw145}

\bibitem{Brook2012}
C.B. {Brook}, G.~{Stinson}, B.K. {Gibson}, J.~{Wadsley}, T.~{Quinn}, \mnras
  \textbf{424}, 1275 (2012).
\newblock \doi{10.1111/j.1365-2966.2012.21306.x}

\bibitem{Aumer2013}
M.~{Aumer}, S.D.M. {White}, T.~{Naab}, C.~{Scannapieco}, \mnras \textbf{434},
  3142 (2013).
\newblock \doi{10.1093/mnras/stt1230}

\bibitem{Vogelsberger2013b}
M.~{Vogelsberger}, S.~{Genel}, D.~{Sijacki}, P.~{Torrey}, V.~{Springel},
  L.~{Hernquist}, \mnras \textbf{436}, 3031 (2013).
\newblock \doi{10.1093/mnras/stt1789}

\bibitem{saitoh08}
T.R. {Saitoh}, H.~{Daisaka}, E.~{Kokubo}, J.~{Makino}, T.~{Okamoto},
  K.~{Tomisaka}, K.~{Wada}, N.~{Yoshida}, \pasj \textbf{60}, 667 (2008)

\bibitem{Hopkins2011}
P.F. {Hopkins}, E.~{Quataert}, N.~{Murray}, \mnras \textbf{417}, 950 (2011).
\newblock \doi{10.1111/j.1365-2966.2011.19306.x}

\bibitem{Hopkins2013c}
P.F. {Hopkins}, D.~{Narayanan}, N.~{Murray}, \mnras \textbf{432}, 2647 (2013).
\newblock \doi{10.1093/mnras/stt723}

\bibitem{Christensen2014b}
C.R. {Christensen}, F.~{Governato}, T.~{Quinn}, A.M. {Brooks}, S.~{Shen},
  J.~{McCleary}, D.B. {Fisher}, J.~{Wadsley}, \mnras \textbf{440}, 2843 (2014).
\newblock \doi{10.1093/mnras/stu399}

\bibitem{Benincasa2016}
S.M. {Benincasa}, J.~{Wadsley}, H.M.P. {Couchman}, B.W. {Keller}, \mnras
  \textbf{462}, 3053 (2016).
\newblock \doi{10.1093/mnras/stw1741}

\bibitem{FIRE2}
P.F. {Hopkins}, A.~{Wetzel}, D.~{Keres}, C.A. {Faucher-Giguere}, E.~{Quataert},
  M.~{Boylan-Kolchin}, N.~{Murray}, C.C. {Hayward}, S.~{Garrison-Kimmel},
  C.~{Hummels}, R.~{Feldmann}, P.~{Torrey}, X.~{Ma}, D.~{Angles-Alcazar}, K.Y.
  {Su}, M.~{Orr}, D.~{Schmitz}, I.~{Escala}, R.~{Sanderson}, M.Y. {Grudic},
  Z.~{Hafen}, J.H. {Kim}, A.~{Fitts}, J.S. {Bullock}, C.~{Wheeler}, T.K.
  {Chan}, O.D. {Elbert}, D.~{Narananan}, ArXiv e-prints  (2017)

\bibitem{Pallottini2017}
A.~{Pallottini}, A.~{Ferrara}, S.~{Bovino}, L.~{Vallini}, S.~{Gallerani},
  R.~{Maiolino}, S.~{Salvadori}, \mnras \textbf{471}, 4128 (2017).
\newblock \doi{10.1093/mnras/stx1792}

\bibitem{Agertz2016}
O.~{Agertz}, A.V. {Kravtsov}, \apj \textbf{824}, 79 (2016).
\newblock \doi{10.3847/0004-637X/824/2/79}

\bibitem{Semenov2016}
V.A. {Semenov}, A.V. {Kravtsov}, N.Y. {Gnedin}, \apj \textbf{826}, 200 (2016).
\newblock \doi{10.3847/0004-637X/826/2/200}

\bibitem{Semenov2018}
V.A. {Semenov}, A.V. {Kravtsov}, N.Y. {Gnedin}, \apj \textbf{861}, 4 (2018).
\newblock \doi{10.3847/1538-4357/aac6eb}

\bibitem{Kormendy2010}
J.~Kormendy, N.~Drory, R.~Bender, M.E. Cornell, \apj \textbf{723}(1), 54
  (2010).
\newblock \doi{10.1088/0004-637X/723/1/54}.
\newblock \urlprefix\url{http://adsabs.harvard.edu/abs/2010ApJ...723...54K}

\bibitem{Shen2010}
J.~Shen, R.M. Rich, J.~Kormendy, C.D. Howard, R.~{De Propris}, A.~Kunder, \apj
  \textbf{720}(1), L72 (2010).
\newblock \doi{10.1088/2041-8205/720/1/L72}.
\newblock \urlprefix\url{http://adsabs.harvard.edu/abs/2010ApJ...720L..72S}

\bibitem{Fisher2011}
D.B. Fisher, N.~Drory, \apj \textbf{733}(2), L47 (2011).
\newblock \doi{10.1088/2041-8205/733/2/L47}.
\newblock \urlprefix\url{http://arxiv.org/abs/1104.0020
  http://stacks.iop.org/2041-8205/733/i=2/a=L47?key=crossref.7e25a6051bda6f1c725b92db72faba5d}

\bibitem{Aumer2014}
M.~Aumer, S.D.M. White, T.~Naab, \mnras \textbf{441}(4), 3679 (2014).
\newblock \doi{10.1093/mnras/stu818}.
\newblock \urlprefix\url{http://arxiv.org/abs/1404.6926
  http://mnras.oxfordjournals.org/cgi/doi/10.1093/mnras/stu818}

\bibitem{Brooks2016}
A.~{Brooks}, C.~{Christensen}, in \emph{Galactic Bulges}, \emph{Astrophysics
  and Space Science Library}, vol. 418, ed. by E.~{Laurikainen}, R.~{Peletier},
  D.~{Gadotti} (2016), \emph{Astrophysics and Space Science Library}, vol. 418,
  p. 317.
\newblock \doi{10.1007/978-3-319-19378-6_12}

\bibitem{SantosSantos2018}
I.M. {Santos-Santos}, A.~{Di Cintio}, C.B. {Brook}, A.~{Macci{\`o}},
  A.~{Dutton}, R.~{Dom{\'{\i}}nguez-Tenreiro}, \mnras \textbf{473}, 4392
  (2018).
\newblock \doi{10.1093/mnras/stx2660}

\bibitem{Garrison-Kimmel2018}
S.~{Garrison-Kimmel}, P.F. {Hopkins}, A.~{Wetzel}, J.S. {Bullock},
  M.~{Boylan-Kolchin}, D.~{Keres}, C.A. {Faucher-Giguere}, K.~{El-Badry},
  A.~{Lamberts}, E.~{Quataert}, R.~{Sanderson}, ArXiv e-prints  (2018)

\bibitem{ElBadry2016}
K.~{El-Badry}, A.~{Wetzel}, M.~{Geha}, P.F. {Hopkins}, D.~{Kere{\v s}}, T.K.
  {Chan}, C.A. {Faucher-Gigu{\`e}re}, \apj \textbf{820}, 131 (2016).
\newblock \doi{10.3847/0004-637X/820/2/131}

\bibitem{Maccio2010}
A.V. {Macci{\`o}}, F.~{Fontanot}, \mnras \textbf{404}, L16 (2010).
\newblock \doi{10.1111/j.1745-3933.2010.00825.x}

\bibitem{Polisensky2011}
E.~{Polisensky}, M.~{Ricotti}, \prd \textbf{83}(4), 043506 (2011).
\newblock \doi{10.1103/PhysRevD.83.043506}

\bibitem{Anderhalden2013}
D.~{Anderhalden}, J.~{Diemand}, \jcap \textbf{4}, 009 (2013).
\newblock \doi{10.1088/1475-7516/2013/04/009}

\bibitem{Horiuchi:2013noa}
S.~Horiuchi, P.J. Humphrey, J.~Onorbe, K.N. Abazajian, M.~Kaplinghat,
  S.~Garrison-Kimmel, Phys. Rev. \textbf{D89}(2), 025017 (2014).
\newblock \doi{10.1103/PhysRevD.89.025017}

\bibitem{Viel2006}
M.~{Viel}, J.~{Lesgourgues}, M.G. {Haehnelt}, S.~{Matarrese}, A.~{Riotto},
  Physical Review Letters \textbf{97}(7), 071301 (2006).
\newblock \doi{10.1103/PhysRevLett.97.071301}

\bibitem{Seljak2006}
U.~{Seljak}, A.~{Makarov}, P.~{McDonald}, H.~{Trac}, Physical Review Letters
  \textbf{97}(19), 191303 (2006).
\newblock \doi{10.1103/PhysRevLett.97.191303}

\bibitem{Viel2008}
M.~{Viel}, G.D. {Becker}, J.S. {Bolton}, M.G. {Haehnelt}, M.~{Rauch}, W.L.W.
  {Sargent}, Physical Review Letters \textbf{100}(4), 041304 (2008).
\newblock \doi{10.1103/PhysRevLett.100.041304}

\bibitem{Boyarsky:2014jta}
A.~Boyarsky, O.~Ruchayskiy, D.~Iakubovskyi, J.~Franse, Phys. Rev. Lett.
  \textbf{113}, 251301 (2014).
\newblock \doi{10.1103/PhysRevLett.113.251301}

\bibitem{Bulbul:2014sua}
E.~Bulbul, M.~Markevitch, A.~Foster, R.K. Smith, M.~Loewenstein, S.W. Randall,
  Astrophys. J. \textbf{789}, 13 (2014).
\newblock \doi{10.1088/0004-637X/789/1/13}

\bibitem{Abazajian:2014gza}
K.N. Abazajian, Phys. Rev. Lett. \textbf{112}(16), 161303 (2014).
\newblock \doi{10.1103/PhysRevLett.112.161303}

\bibitem{Barkana2001}
R.~{Barkana}, Z.~{Haiman}, J.P. {Ostriker}, \apj \textbf{558}, 482 (2001).
\newblock \doi{10.1086/322393}

\bibitem{Mesinger2005}
A.~{Mesinger}, R.~{Perna}, Z.~{Haiman}, \apj \textbf{623}, 1 (2005).
\newblock \doi{10.1086/428770}

\bibitem{deSouza2013}
R.S. {de Souza}, A.~{Mesinger}, A.~{Ferrara}, Z.~{Haiman}, R.~{Perna},
  N.~{Yoshida}, \mnras \textbf{432}, 3218 (2013).
\newblock \doi{10.1093/mnras/stt674}

\bibitem{Pacucci2013}
F.~{Pacucci}, A.~{Mesinger}, Z.~{Haiman}, \mnras \textbf{435}, L53 (2013).
\newblock \doi{10.1093/mnrasl/slt093}

\bibitem{Governato2015}
F.~{Governato}, D.~{Weisz}, A.~{Pontzen}, S.~{Loebman}, D.~{Reed}, A.M.
  {Brooks}, P.~{Behroozi}, C.~{Christensen}, P.~{Madau}, L.~{Mayer}, S.~{Shen},
  M.~{Walker}, T.~{Quinn}, B.W. {Keller}, J.~{Wadsley}, \mnras \textbf{448},
  792 (2015).
\newblock \doi{10.1093/mnras/stu2720}

\bibitem{Chau2017}
A.~{Chau}, L.~{Mayer}, F.~{Governato}, \apj \textbf{845}, 17 (2017).
\newblock \doi{10.3847/1538-4357/aa7e74}

\bibitem{Spergel:1999mh}
D.N. Spergel, P.J. Steinhardt, Phys.Rev.Lett. \textbf{84}, 3760 (2000).
\newblock \doi{10.1103/PhysRevLett.84.3760}

\bibitem{Yoshida2000}
N.~{Yoshida}, V.~{Springel}, S.D.M. {White}, G.~{Tormen}, \apjl \textbf{544},
  L87 (2000).
\newblock \doi{10.1086/317306}

\bibitem{Miralda2002}
J.~{Miralda-Escud{\'e}}, \apj \textbf{564}, 60 (2002).
\newblock \doi{10.1086/324138}

\bibitem{Peter:2012jh}
A.H.G. Peter, M.~Rocha, J.S. Bullock, M.~Kaplinghat, Mon. Not. Roy. Astron.
  Soc. \textbf{430}, 105 (2013).
\newblock \doi{10.1093/mnras/sts535}

\bibitem{Loeb:2010gj}
A.~Loeb, N.~Weiner, Phys. Rev. Lett. \textbf{106}, 171302 (2011).
\newblock \doi{10.1103/PhysRevLett.106.171302}

\bibitem{Elbert:2014bma}
O.D. Elbert, J.S. Bullock, S.~Garrison-Kimmel, M.~Rocha, J.~Oñorbe, A.H.G.
  Peter, Mon. Not. Roy. Astron. Soc. \textbf{453}(1), 29 (2015).
\newblock \doi{10.1093/mnras/stv1470}

\bibitem{Fry2015}
A.B. {Fry}, F.~{Governato}, A.~{Pontzen}, T.~{Quinn}, M.~{Tremmel},
  L.~{Anderson}, H.~{Menon}, A.M. {Brooks}, J.~{Wadsley}, \mnras \textbf{452},
  1468 (2015).
\newblock \doi{10.1093/mnras/stv1330}

\bibitem{Vogelsberger2014}
M.~{Vogelsberger}, J.~{Zavala}, C.~{Simpson}, A.~{Jenkins}, \mnras
  \textbf{444}, 3684 (2014).
\newblock \doi{10.1093/mnras/stu1713}

\bibitem{Dooley2016}
G.A. {Dooley}, A.H.G. {Peter}, M.~{Vogelsberger}, J.~{Zavala}, A.~{Frebel},
  \mnras \textbf{461}, 710 (2016).
\newblock \doi{10.1093/mnras/stw1309}

\bibitem{Breddels2013}
M.A. {Breddels}, A.~{Helmi}, R.C.E. {van den Bosch}, G.~{van de Ven},
  G.~{Battaglia}, \mnras \textbf{433}, 3173 (2013).
\newblock \doi{10.1093/mnras/stt956}

\bibitem{Walker2011}
M.G. {Walker}, J.~{Pe{\~n}arrubia}, \apj \textbf{742}, 20 (2011).
\newblock \doi{10.1088/0004-637X/742/1/20}

\bibitem{Strigari2014}
L.E. {Strigari}, C.S. {Frenk}, S.D.M. {White}, ArXiv e-prints  (2014)

\end{thebibliography}

%
\chapter*{Participants}
The Simons Symposium on ``Illuminating Dark Matter'' took place on May 13-19, 2018 at Schloss Elmau, in Elmau, Kr\"un, Germany.  

\vskip 1cm
\begin{tabular}{@{}p{5cm}@{}p{7.2cm}@{}}

Kevork Abazajian &  UC Irvine  \\
Jo Bovy &	University of Toronto  \\
Alyson Brooks & Rutgers University  \\
Bernard Carr & Queen Mary, University of London  \\
Aaron Chou & Fermilab  \\
Neal Dalal & Perimeter Institute  \\
Bertrand Echenard & California Institute of Technology  \\
Rouven Essig & Stony Brook University  \\
Jonathan L.~Feng & UC Irvine  \\
Roni Harnik & Fermilab  \\
Phil Hopkins  &California Institute of Technology  \\
Manoj Kaplinghat  &UC Irvine  \\
Alexander Kusenko  &UCLA and Kavli IPMU  \\
Rafael F.~Lang & Purdue University  \\
Julio Navarro  &University of Victoria  \\
Mauro Raggi & Rome La Sapienza  \\
Adam Ritz & University of Victoria  \\
Joshua Ruderman & New York University  \\
Jessie Shelton & University of Illinois at Urbana-Champaign \\
Javier Tiffenberg & Fermi National Accelerator Laboratory  \\
Tomer Volansky & Tel Aviv University  \\
Neal Weiner & New York University  \\
Kathryn Zurek & Lawrence Berkeley National Laboratory  
\end{tabular}

\end{document}